\documentclass[preprint]{aastex62}

\usepackage{amsmath}
\usepackage{amssymb}
\usepackage[normalem]{ulem}

\newcommand{\bff}{\boldsymbol{f}}
\newcommand{\by}{\boldsymbol{y}}

\newcommand{\bX}{\boldsymbol{X}}
\newcommand{\bbeta}{\boldsymbol{\beta}}
\newcommand{\bepsilon}{\boldsymbol{\epsilon}}
\newcommand{\bLambda}{\boldsymbol{\Lambda}}
\newcommand{\bSigma}{\boldsymbol{\Sigma}}

\usepackage{natbib}

\received{January 25, 2019}
\revised{April 12, 2019}
\accepted{?}
\submitjournal{PASP}

\shorttitle{Multivariate time series analysis}
\shortauthors{Eyer et al.}

\begin{document}

\title{Multivariate time series analysis of variable objects in the \textit{Gaia} mission}

\correspondingauthor{Laurent Eyer}
\email{Laurent.Eyer@unige.ch}

\author{Laurent Eyer}
\affil{University of Geneva,
Chemin des Maillettes 51,
CH-1290 Versoix, Switzerland}

\author{Maria S\"uveges}
\affiliation{University of Geneva, Chemin d'Ecogia 16, CH-1290 Versoix, Switzerland}
\affiliation{Max Planck Institute for Astronomy, K\"onigstuhl 17, D-69117 Heidelberg, Germany}

\author{Joris De Ridder}
\affiliation{Instituut voor Sterrenkunde, KU Leuven, Celestijnenlaan 200D, B-3001 Leuven, Belgium}

\author{Sara Regibo}
\affiliation{Instituut voor Sterrenkunde, KU Leuven, Celestijnenlaan 200D, B-3001 Leuven, Belgium}

\author{Nami Mowlavi}
\affiliation{University of Geneva,
Chemin des Maillettes 51,
CH-1290 Versoix, Switzerland}
\affiliation{University of Geneva, Chemin d'Ecogia 16, CH-1290 Versoix, Switzerland}

\author{Berry Holl}
\affiliation{University of Geneva,
Chemin des Maillettes 51,
CH-1290 Versoix, Switzerland}
\affiliation{University of Geneva, Chemin d'Ecogia 16, CH-1290 Versoix, Switzerland}

\author{Lorenzo Rimoldini}
\affiliation{University of Geneva, Chemin d'Ecogia 16, CH-1290 Versoix, Switzerland}

\author{Fran\c cois Bouchy}
\affiliation{University of Geneva,
Chemin des Maillettes 51,
CH-1290 Versoix, Switzerland}




\begin{abstract}
In astronomy, we are witnessing an enormous increase in the number of source detections, precision, and diversity of measurements. Additionally, multi-epoch data is becoming the norm, making time series analyses an important aspect of current-day astronomy.   
The \textit{Gaia} mission is an outstanding example of a multi-epoch survey that provides measurements in a large diversity of domains, with its broad-band photometry, spectrophotometry in blue and red (used to derive astrophysical parameters), spectroscopy (employed to infer radial velocities, $v \sin(i)$, and other astrophysical parameters), and its extremely precise astrometry. Most of all that information is provided for sources covering the entire sky.

    Here, we present several properties related to the \textit{Gaia} time series, such as the time sampling, the different types of measurements, the \textit{Gaia} $G$, $G_{\rm{BP}}$ and $G_{\rm{RP}}$-band photometry, and \textit{Gaia}-inspired studies using the CORAVEL (CORrelation-RAdial-VELocities) data to assess the potential of the information on the radial velocity, the FWHM, and the contrast of the cross-correlation function. We also present techniques (which are used or are under development) that optimise the extraction of astrophysical information from the different instruments of \textit{Gaia}, such as the principal component analysis and the multi-response regression.
The detailed understanding of the behaviour of the observed phenomena in the various measurement domains can lead to richer and more precise characterization of the \textit{Gaia} data, including the definition of more informative attributes that serve as input to (our) machine learning algorithms.

\end{abstract}



~\\

\section{Introduction} \label{sec:intro}

Time series provide essential information for understanding our Universe and its constituents. 
The measured quantities can be of different nature: astrometric, photomeric or spectroscopic, and the time domain analysis provides generally an invaluable insight on specific topics, see figure 1 of \citet{Eyer2018}.
Time series can be analysed not only individually but also in combination with other time series, in order to take advantage of the dependence that might exist among them. The \textit{Gaia} data provide a compelling case for the analysis of multi-variate time series and we elaborate on some of such aspects in this article.

The \textit{Gaia} mission \citep{PrustiEtAl2016} is quite unique among the many multi-epoch surveys for several reasons: its extremely precise astrometry, about two orders of magnitude better than the previous state-of-the-art astrometric mission \textit{Hipparcos} \citep{esahipparcos}, the homogeneity and depth of the survey over the entire sky (with four orders of magnitude more sources detected than in \textit{Hipparcos}), and the multiple instruments that observe near simultaneously each object. 

The analysis of the \textit{Gaia} data by the Data Processing and Analysis Consortium (DPAC) is iterative and progressively increasing in complexity. In the context of variable object data releases, this translates into larger number and diversity of variability types of the published sources, as well as greater input data volumes and new data types.
In the \textit{Gaia} first data release (DR1), we published 3,194 RR Lyrae stars and Cepheids (from the region around the South Ecliptic Pole), using only the $G$-band time series \citep{EyerEtAl2017, ClementiniEtAL2016}. In DR2, the variability analysis and processing coordination unit published 550,737 sources distributed over 6 main variability groups, using the (mean) per-field-of-view $G$-band photometry, the per-CCD $G$-band photometry, the integrated $G_{\rm{BP}}$ and $G_{\rm{RP}}$ spectrophotometry, and the per-source astrometry \citep{HollEtAl2018}. For DR3, we intend to publish several millions of variable sources for up to about two dozen variability types.

Further detailed exploitation of the data includes the analysis of time series of the spectra from the Radial Velocity Spectrometer (RVS), which might detect transient emission lines, or highlight variations in the radial velocity estimates or in the spectral line full width at half maximum (FWHM) and contrast of the cross-correlation function (CCF). The time series of the astrometric measurements are used to derive the parallax and proper motion, the orbital parameters for binary systems, but can also be used to identify or validate transient events like microlensing \citep{KlueterEtAl2018,RybickiEtAl2018}.

In the variability analysis for \textit{Gaia}, we combine the various multi-epoch time series (and their derived parameters produced by other data processing centres in the \textit{Gaia} DPAC) into a variety of \textit{attributes} like period, colour, amplitude,  etc., that accurately characterize the variable source signal encoded in the data.

The detection of periodicity and how it is affected by the time sampling are described in Section~\ref{sec:period}, some of the methods to combine multiple (photometric) time series are presented in Section~\ref{sec:multibandphotometry}, the additional value from the analysis of spectrophotometric time series is outlined in Section~\ref{sec:spphot}, the features of radial velocity time series for some variability types are described in Section~\ref{sec:floats}, and conclusions are drawn in Section~\ref{sec:conclusion}.

\begin{figure*}
\begin{center}
\includegraphics[width=18cm]{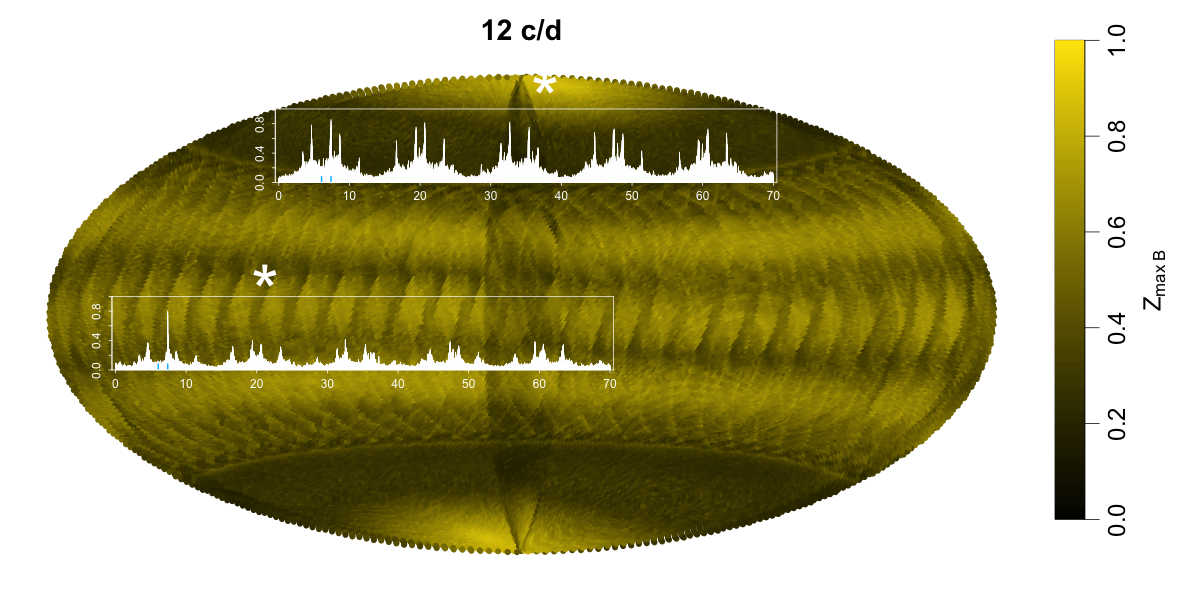}
\caption{Sky map of the alias strength at $n = 3$ (12 d$^{-1}$), colour-coded from black to yellow, and the periodograms of a simulated $\delta$\,Scuti-type star after the commissioning phase and a subsequent 5-year period of \textit{Gaia} nominal scanning law, if situated in a high-alias region near the north ecliptic pole (top inset in white) or in a low-alias region near the ecliptic plane (bottom inset). The $x$-axis of the insets shows frequencies in units of cycles per day, the $y$-axis shows the relative gain in $\chi^2$ of fitting a sinusoidal as compared to fitting a constant. The two pulsation frequencies with the highest amplitudes are indicated by the blue tickmarks at the bottom of the periodograms. The precise locations of the simulated $\delta$ Scuti stars are indicated by white stars. The background shows the relative strength of one of the highest typical \textit{Gaia} aliasing frequency over the sky, namely, the one at 12~d$^{-1}$. Aliasing is minimal in dark regions, and the highest in yellow-coloured ones.
}
\label{fig:DSCTsim}       
\end{center}
\end{figure*}

\section{Period and time sampling} \label{sec:period}

One of the most informative attributes to distinguish between different variability types is the dominant timescale of the variations. Many variable stars show more or less strictly  repetitive changes in their brightness, colour, or spectrum, with specific period(s), whose range is characteristic to the variability type. A reliable estimate of periods is therefore crucial for the quality of classification and for the analysis of variable stars. 

As it happens for many wide-field time-domain surveys, \textit{Gaia}'s  peculiar sparse time sampling complicates the period search, though the irregularity of the sampling pushes the Nyquist frequency higher \citep{EyerBartholdi1999}. Due to its position at the L2 point following the Earth's orbit around the Sun, to its rotation and precession, and to its two telescopes, the spacecraft scans each point on the sky multiple times during the mission timespan in a peculiar sequence, which is called the Nominal Scanning Law (NSL) \citep[see][for some general characteristics of the sampling and of the spectral windows]{EyerEtAl2017}. These cadences are different at each sky location and produce some difficulties for period search using the \textit{Gaia} time series: even for an object with a single strictly periodic signal, the periodogram (the goodness-of-fit statistics of a light curve fit, most often a sine-wave, as a function of the frequency) shows a series of humps, bumps and sharp peaks. The apparent `leaking' of the signal to other frequencies, distant from the true frequency, is the consequence of the sparse, quasi-periodically clustered sampling of the signal, and can lead to the well known phenomenon of `aliasing' in astronomy. 

These alias peaks appear at characteristic intervals from the true frequency in the periodogram, determined by any quasi-periodicity in the time sampling of the observations. If a signal of frequency $f_s$ is observed quasi-regularly with a frequency $f_o$, the periodogram consists of a series of peaks at $| f_s \pm n f_o |$, where $n$ is a small integer. For \textit{Gaia}, the most important sampling frequency corresponds to its rotation: $f_o \approx 4$\,d$^{-1}$. But the complexity of its motion dramatically modifies this regular pattern and induces an even more complex alias structure, which renders the identification of the correct period from the periodograms based on some \textit{Gaia} observing sequences challenging.

Figure~\ref{fig:DSCTsim} illustrates samples of such intricated aliasing structures. The background, colour-coded from black to yellow, shows the typical amplitudes of one of the most characteristic \textit{Gaia} alias peaks (at $n = 3$, that is, at 12\,d$^{-1}$), relative to the peak at the true frequency in a noiseless case. Black/dark colour indicates areas where the sampling does not produce high aliasing (the relative size of the alias peaks is close to zero), whereas in the yellow regions, aliasing is expected to be high (close to one, implying alias peaks nearly as high as the correct peak even in a noiseless case). Fig.~\ref{fig:DSCTsim} suggests that aliasing can vary strongly and on short spatial scales over the sky. 

The two periodograms superposed to the alias map in Fig.~\ref{fig:DSCTsim} demonstrate the consequences of the aliasing on the detection and identification of frequencies. We simulated a $\delta$\,Scuti star with five oscillation modes, modelled after X\,Caeli \citep{mantegazzaporetti92}. We sampled it according to two different cadences of \textit{Gaia} corresponding to two different sky locations (both with about 70 field-of-view observations), added noise corresponding to a 19 magnitude star observed by \textit{Gaia} in the $G$ band, and computed the periodograms. The time sampling has an obvious strong effect on the detectability of the oscillation modes of the signal: whereas in the low-aliasing region, the strongest oscillation (with an amplitude of 36.9 mmag) clearly stands out, in the high-aliasing regions there are many peaks of comparable sizes, and an alias peak is identified as the main pulsation frequency. The remaining four oscillations, with amplitudes of 7.5 mmag or lower, do not appear in the periodogram, neither aliased, nor at their correct frequencies. 

The problem of aliasing, in conjunction with the importance of the variability periods, is one of the motivations for the search for the most efficient use of all the information contained in the variety of \textit{Gaia} data types.

\section{Multi-band photometry} \label{sec:multibandphotometry}

As shown in fig.~11 of \cite{EyerEtal2018}, the different physical origins of light variations (due to pulsation, eruption, rotation, or eclipses) can be distinguished from their patterns drawn in the colour-magnitude diagram as a function of time. 
The \textit{Gaia} quasi-simultaneous independent measurements in several bands help characterize the variability with important clues on its origin, which improves the ability to recover the correct period (when the objects have a periodic behaviour).


\subsection{Principal Component Analysis} \label{subsec:pca}

\begin{figure*}
\begin{center}
\includegraphics[width=8.9cm]{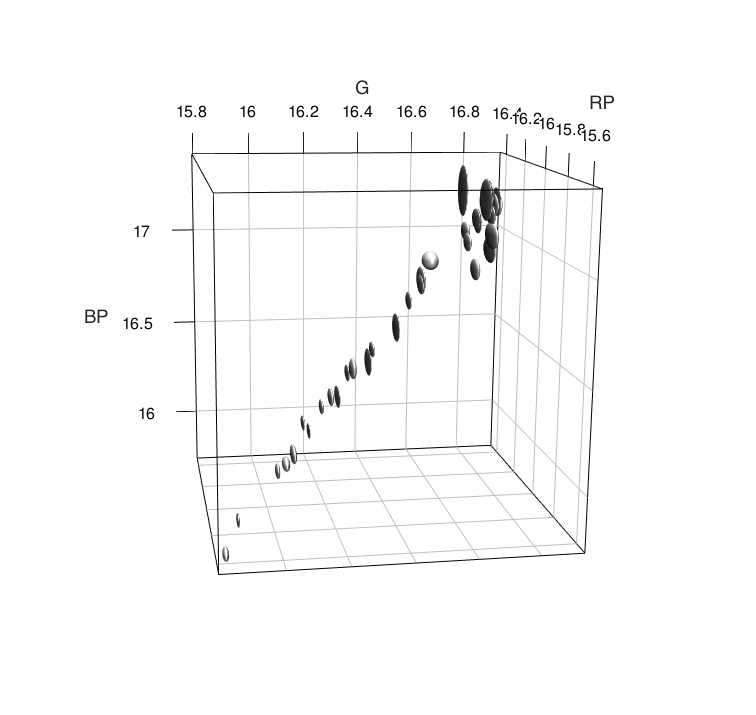}
\includegraphics[width=8.9cm]{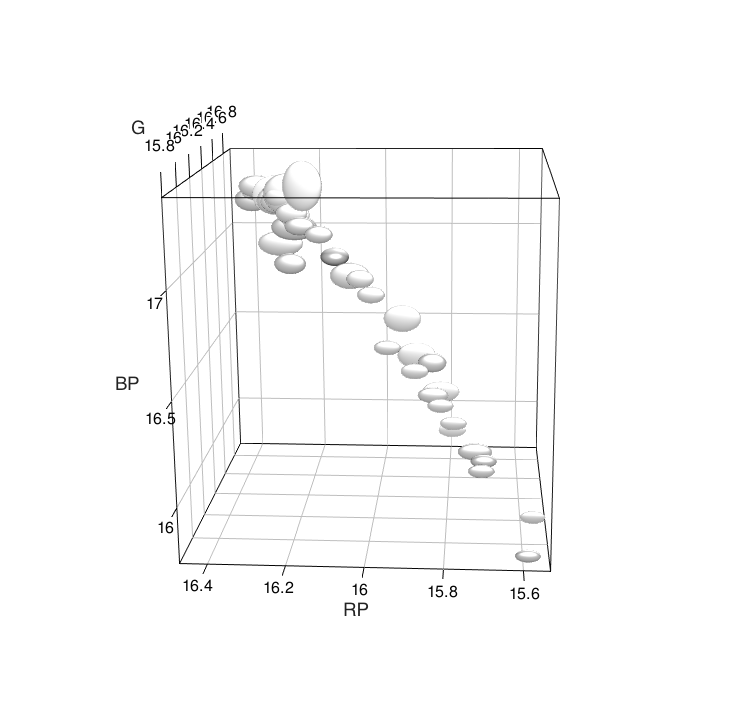}
\caption{Two views of the multi-band photometry of an RR\,Lyrae star as observed by  \textit{Gaia} in the $G$, $G_{\rm{BP}}$, and $G_{\rm{RP}}$ filters. Each ellipsoid corresponds to one set of near-simultaneous observations in the three bands, with sizes along the three coordinate axes corresponding to the nominal standard errors reported by the \textit{Gaia}~DR2 database.  
}
\label{fig:3dplot}       
\end{center}
\end{figure*}

The high correlation between the near-simultaneous photometric measurements in different bands, as presented for an RR Lyrae-type variable star in Fig.~\ref{fig:3dplot}, suggests a first idea to exploit the information provided by multi-band data. It is evident that the variability amplitude is the largest not in any of the individual bands, but along a specific direction in the space spanned by the different bands. Thus, variability detection by looking for the excess scatter along this direction can attain more favourable detection limits than in any of the original bands. Period search also can be expected to provide better results, thanks to the improved signal-to-noise ratio (SNR). Moreover, the geometry and the tilt of the point cloud in the 3-dimensional space can be characteristic of the variability type, reflecting the dissimilar colour variations of eclipsing binaries versus pulsating or spotted stars, and also the average colour differences between pulsating types.

Principal component analysis \citep[PCA; e.g.,][]{jolliffe,hastieetal} identifies the direction of maximal variability via a singular value decomposition. Let $\mathbf{X}$ denote the multi-band photometric time series of an object, arranged into matrix form with $M$ columns, each representing the time series of a single band, and $N$ rows, each representing a vector of simultaneous photometry at one observing epoch. Then PCA is equivalent to 
$$
\mathbf{X}^T \mathbf{X} = \mathbf{V} \mathbf{D}^2 \mathbf{V}^T,
$$
where $\mathbf{X}^T \mathbf{X}$ is the sample covariance matrix multiplied by the number of observations (the superscript $T$ denotes transposition); the columns of $\mathbf{V}$ are the eigenvectors, that is, the principal directions of $\mathbf{X}$; and $\mathbf{D}^2$ is the diagonal matrix of eigenvalues, with non-negative entries $d_1^2, \ldots, d_M^2$, which are the variances of the projections of the observations to the principal direction. By convention, $d_1^2 \geq \ldots \geq d_M^2$, so the first principal direction corresponds to the direction of the maximal variance. 

If the variations in the different bands are approximately synchronized, the point cloud in the 3-dimensional space forms a narrow, elongated ellipsoid, as seen in the left panel of Fig.~\ref{fig:3dplot} for an RR Lyrae star. The time series of the projections of the observations to the direction of the longest axis of this ellipsoid is the PC1 time series, computed as $\mathbf{X} \mathbf{v}_1$, where $\mathbf{v}_1$ is the first column of $\mathbf{V}$. \citet{suvegesetal12b} showed on the Sloan Digital Sky Survey (SDSS) Stripe 82 data that this can be used for variability detection and for period search in a similar way as the individual band-wise time series, with better results. In particular, performing the period search on PC1 time series for a known RR\,Lyrae sample from the SDSS Stripe 82 resulted in fewer aliased period detections than a similar period search performed on the $g$-band (which is usually the least noisy among the five SDSS bands). Furthermore, the PC1 direction (summarized by the coefficients of the different bands in the linear combination defining the PC1) turned out to characterize whether the colour of the star varies or not. This enables a distinction between RRc-type pulsating variables and contact eclipsing binaries, which have similar sinusoidal light curves, but RRc stars do have colour variation over their pulsation cycle, whereas eclipsing binaries do not. The PCA coefficients can thus be used as classification attributes, together with the (improved) period based on PC1.

Figure~\ref{fig:3dplot} indicates also some differences between \textit{Gaia} and SDSS Stripe 82 data, namely, the higher level of noise in two of the \textit{Gaia} bands as compared to the third. The \textit{Gaia} $G_{\rm{BP}}$ and $G_{\rm{RP}}$ photometries typically have a lower SNR than the $G$ photometry, which reduces the potential for improvement expected from a combination. 
The utility of PCA on \textit{Gaia} data depends strongly on various characteristics of the object such as its apparent brightness, spectrum, colour, sky position, and observing configuration. Investigations concerning the manner and the conditions under which PCA can be optimally exploited for \textit{Gaia} data are underway. Since the period is a key piece of information about the variability type, alternative methods for period search using the multi-band information are also being explored, as presented in the next section. 

\subsection{Multi-Response Regression}
The classical least-squares regression of a single time series can be easily generalized to what is usually
called \textit{multi-response regression} which fits a model to a vector of responses. In the case of \textit{Gaia},
the different responses are the different time series of $G$, $G_{\rm{BP}}$, and $G_{\rm{RP}}$ photometry of the same star. 
There are several ways to write down the multi-response model, but here we opt for the "rolled-out" form where 
the responses 
$\by_j = (y_{j,0}, \cdots, y_{j,N})^T$ are listed in single large vector 
$\by \equiv (\by_1^T, \by_2^T, \cdots, \by_D^T)$ with $D=3$, and for which we write
\begin{equation}
\by = \bff(\bbeta) + \bepsilon,
\end{equation}
where $\bff(\bbeta)$ is the model with model parameters $\bbeta$, and $\bepsilon$ is the noise vector 
(Seber \& Wild, 1989). This form was also used by Vanderplas \& Ivezi\'c (2015), and conveniently allows 
for different time points and a different model for each of the responses. Contrary to the latter authors, we opt to do the harmonic regression in a complex Hilbert space using the model
\begin{equation}
\label{harmonicmodel}
f_j(t \ | \ \omega) \equiv \beta_{0} + \sum_{n=1}^{M} \beta_{n} \ e^{i n \omega t} + \beta_{j,0} + \sum_{n=1}^{ M} \beta_{j,n} \ e^{i n \omega t},
\end{equation}
where $t$ denotes the time, $\omega$ the angular frequency, and $M$ is the number of harmonics for each model which is considered an input parameter. Note that the model is linear in its coefficients $\beta_i \in \mathbb{C}$. The first part is a base model common to all passbands, while the second part characterizes the deviation from this base model for each passband in particular.
The advantage of working in complex space is a more convenient way to update the design matrix for each frequency (in double precision), when using a fixed frequency step $\Delta\omega$. 
Given the $k$-th frequency $\omega_k = \omega_0 + k \Delta\omega$ of the frequency spectrum, we can define the design matrix
\begin{equation}
\bX^{(k)} \equiv 
\begin{bmatrix}
1 & e^{i \omega_k t_0} & \cdots & e^{i M \omega_k t_0} \\
1 & e^{i \omega_k t_1} & \cdots & e^{i M \omega_k t_1} \\
\vdots & \vdots & \cdots & \vdots \\
1 & e^{i \omega_k t_N} & \cdots & e^{i M \omega_k t_N} 
\end{bmatrix}
\end{equation}
and its increment
\begin{equation}
    \bX^{\delta} \equiv 
    \begin{bmatrix}
    1 & e^{i \Delta\omega t_0} & \cdots & e^{i M \Delta\omega t_0} \\
    1 & e^{i \Delta\omega t_1} & \cdots & e^{i M \Delta\omega t_1} \\
    \vdots & \vdots & \cdots & \vdots \\
    1 & e^{i \Delta\omega t_N} & \cdots & e^{i M \Delta\omega t_N} 
    \end{bmatrix}
\end{equation}
so that the update equation is 
\begin{equation}
\bX^{(k+1)} = \bX^{(k)} \circ \bX^{\delta}
\end{equation}
where $\circ$ denotes the Hadamard (or elementwise) product. This avoids recomputing all the sines and cosines
for every frequency of the frequency spectrum, and speeds up the computations significantly (by about an order of magnitude for \textit{Gaia}-like $G$, $G_{\rm{BP}}$, and $G_{\rm{RP}}$ time series).

The complex regression then comes down to minimizing the function
\begin{equation}
S(\omega_k) \equiv (\by - \bX^{(k)} \bbeta)^* \ \bSigma^{-1} \ (\by - \bX^{(k)} \bbeta) + \bbeta^* \bLambda \bbeta
\end{equation}
where $*$ denotes the conjugate transpose and $\bSigma$ is the covariance matrix of the observations $\by$. The last part $\bbeta^* \bLambda \bbeta$ is called $L2$ or sometimes ridge regularization which is required as there is a degeneracy between the parameters coming from the common base model and those of the passband-specific model. For a heuristic choice for a proper $\Lambda$ we refer to Vanderplas \& Ivezi\'c (2015). We currently approximate $\bSigma$ by a diagonal matrix containing the variances of the observations, but neglecting any possible inter- and intra-correlations (i.e., off-diagonal elements). 

We define the frequency spectrum $P(\omega)$ in the same way as in Vanderplas \& Ivezi\'c (2015):
\begin{equation}
P(\omega) = \frac{\by^T \bSigma^{-1} \bX \left( \bX^{*} \bSigma^{-1} \bX + \bLambda \right)^{-1} \bX^{*} \bSigma^{-1} \by}{\by^T \bSigma^{-1} \by}.
\end{equation}

\section{Spectrophotometry} \label{sec:spphot}
\begin{figure*}
\begin{center}
\includegraphics[width=8.9cm]{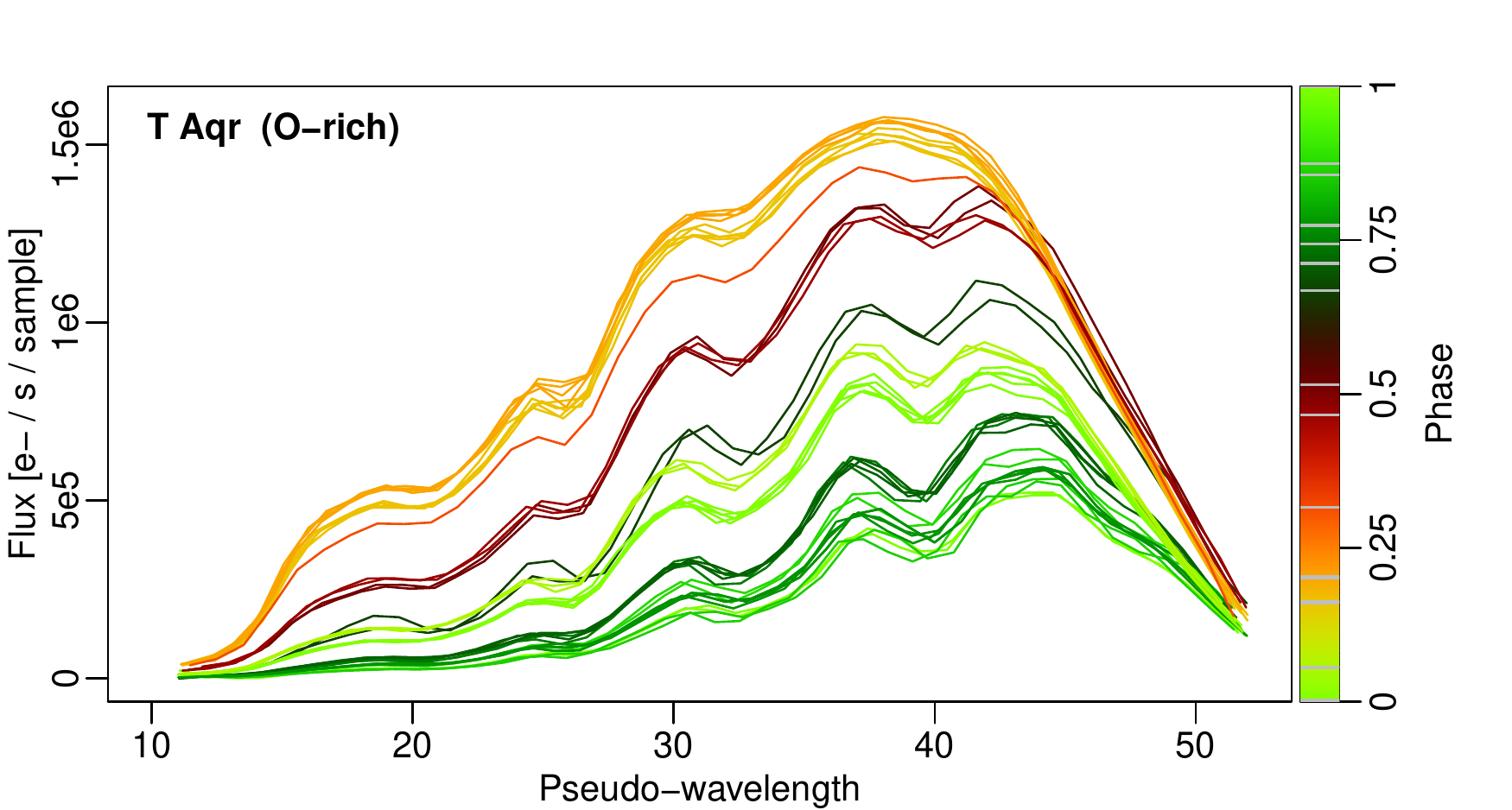}
\includegraphics[width=8.9cm]{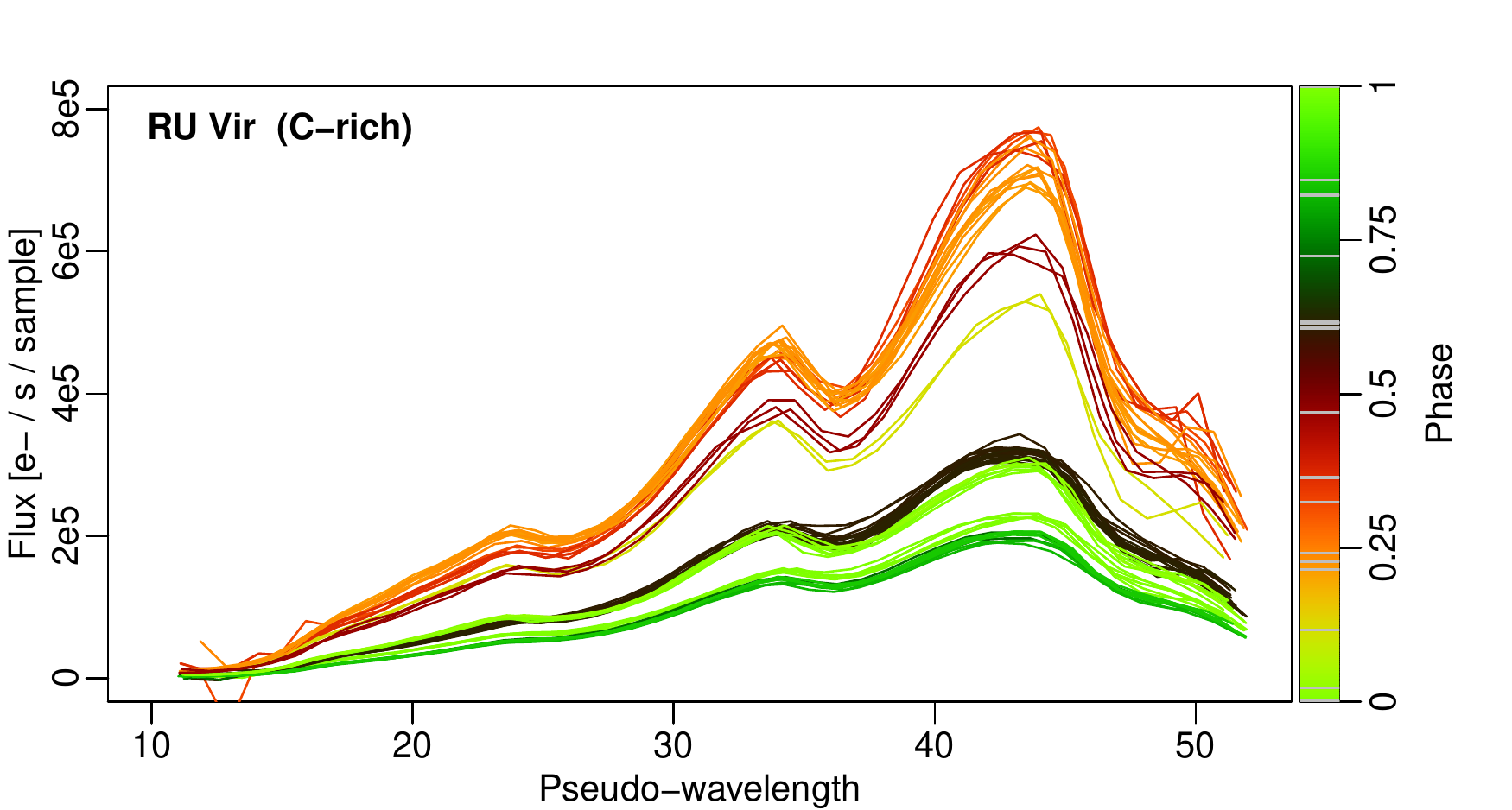}
\caption{The time series of the epoch spectra of the O-rich long period variable T\,Aqr (left) and the C-rich long period variable RU\,Vir (right). The spectra are shown as a function of pseudo-wavelengths, covering approximately the wavelength interval between 640 and 1050 nm. The flux unit is photo-electron number per second per pixel. The spectra are colour-coded by phase in the pulsation cycle; the colour legend is shown on the right-hand side of each panel, with light grey segments indicating the phases of the individual $G_{\rm{RP}}$ transits. (From the \textit{Gaia} Image of the week of 15/11/2018, by ESA/Gaia/DPAC, Mowlavi et al.) 
}
\label{fig:iow}       
\end{center}
\end{figure*}



In this era of information explosion, surveys are planned or are underway with the aim to gather time-resolved spectroscopic information. An example is the Time-Domain Spectroscopic Survey \citep{ruanetal16,macleodetal18}, a subproject of SDSS eBOSS, which aims to collect spectra about certain types of variable objects. \textit{Gaia} measures low-resolution spectra in blue (330-680~nm) and red (640-1050~nm) at each transit for all the objects down to its detection limit, which provides a goldmine of information for variability detection, classification, and analysis of large samples from the visible part of the Milky Way.

The scientific potential of the time series of \textit{Gaia} $G_{\rm{RP}}$ spectra is illustrated in Fig.~\ref{fig:iow}. The two panels compare the spectra of two long-period variable stars with different chemical signatures. Both are evolved red giant stars with large-amplitude radial-mode pulsations. The spectrum of the O-rich T\,Aqr (on the left) is dominated by the absorption lines of TiO, while that of the C-rich RU\,Vir shows the presence of CH and C$_2$. The signatures of the two compositions are visibly distinct. The absorption features of the C-rich star are wider and fewer than those of the O-rich star, which potentially enables the identification of the rare C-rich red giant stars. 
Moreover, variations in the spectrum and in the absorption features can be followed through the pulsation cycle. The absorption features vary in a different way in the two stars: whereas all the identifiable lines are the strongest at maximum brightness for RU\,Vir, the features behave in a more complex way for T\,Aqr, due to the temperature sensitivity of the lines.

The use of the time series of spectra is therefore very promising, both for various tasks associated with the \textit{Gaia} variable star catalog production and for science in general, especially if we consider the size of the \textit{Gaia} sample and its unprecedented representation of Galactic structures. The statistical literature offers a wide range of methodologies for functional time series analysis, where instead of the classical real-valued random variable (the photometry of the star), a function (spectrum or spectral line) is considered the variable of interest as a function of time. Investigations have just started, most importantly, with the application of the well-known and broadly applied functional PCA to separate elements of variations in the spectra, but other methods ranging from the derivation of ad~hoc classification attributes to more complex descriptions of the functional time series structure will also be considered.

\section{Radial velocities and photometry} \label{sec:floats}



We performed a preliminary study of the photometry and CCF of a spectrometer by investigating a subset of \textit{Hipparcos} variable stars that were observed by the CORAVEL spectrometer \citep{1979VA.....23..279B}. The \textit{Hipparcos} periodic star catalogue \citep{Eyer1998} contains 2712 stars. The variability types provided in this catalogue were extracted from the literature and from visual inspection of the phase-folded light curves. \cite{DubathEtAl2011} developed a systematic, automated classification of the complete sample of \textit{Hipparcos} periodic variable stars into 26 variability types using Random Forest \citep{Breiman.Random.Forest}. These authors used a set of objects of known type to train their supervised classification algorithms. They selected a sub-set of 1661 \textit{Hipparcos} stars with the most reliable types. They noticed that the most useful attributes evaluated with the random forest methodology included, in decreasing order of importance, the period, the amplitude, the $V - I$ colour index, the absolute magnitude, the residual around the folded light-curve model, the skewness of the magnitude distribution and the amplitude of the second harmonic of the Fourier series model relative to that of the fundamental frequency. They also noticed that the main misclassification cases, up to a rate of about 10\%, arise due to confusion between slowly pulsating B stars (SPB) and $\alpha^2$\,Canum Venaticorum (ACV) stars, and between eclipsing binaries, ellipsoidal variables and other variability types. 

\begin{figure*}
\begin{center}
\includegraphics[width=0.6\textwidth]{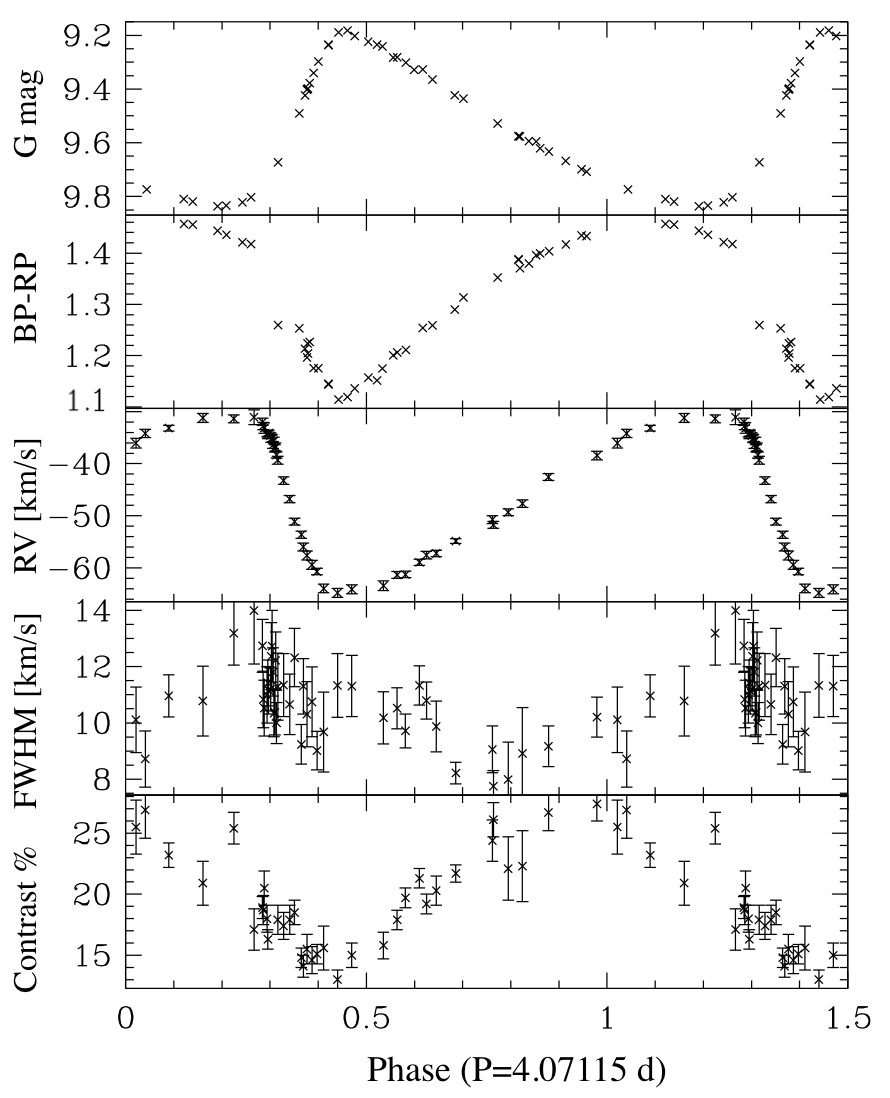}
\caption{The \textit{Gaia}~DR2 photometry ($G$-band and $G_{\rm{BP}}-G_{\rm{RP}}$ data) of a classical Cepheid (HIP 1213), together with the CORAVEL parameters derived from the cross-correlation function (CCF): the radial velocity (RV), the full width at half maximum (FWHM), and the CCF contrast. 
}
\label{fig:CCF}       
\end{center}
\end{figure*}

A total of 160 of these 1661 periodic variable stars were observed more than 10 times with CORAVEL with a precision of about 1 km/s.
The information derived from the CORAVEL epoch spectra included the radial velocity (RV in km\,s$^{-1}$), the CCF contrast (relative depth of the CCF as percentage), and the CCF FWHM (in km\,s$^{-1}$). 
In Fig.~\ref{fig:CCF}, we present these parameters for a pulsating star, a 4-day period Cepheid, for which the \textit{Gaia}~DR2 photometry (the $G$-band data and the $G_{\rm{BP}}-G_{\rm{RP}}$ colour) was published.

Despite the limited number of \textit{Hipparcos} periodic variables that were sufficiently sampled by CORAVEL, automated classification tests confirmed the usefulness of quantities derived from CORAVEL (five of which ranked in the top-10 most useful classification attributes).
The phase and shape of the RVs and CCF parameters can be used in combination with photometric data to improve the period determination and the classification of periodic variables. 
For example, Fig.~\ref{fig:CCF} shows features of pulsating stars that are common to many types: the colour is the the bluest at maximum brightness, while the radial velocity and contrast are close to minimum at maximum brightness.
The time series of full width at half maximum shows variations as well with the same periodicity, although offset in phase with respect to the other quantities.

\begin{figure*}
\begin{center}
\includegraphics[width=0.6\textwidth]{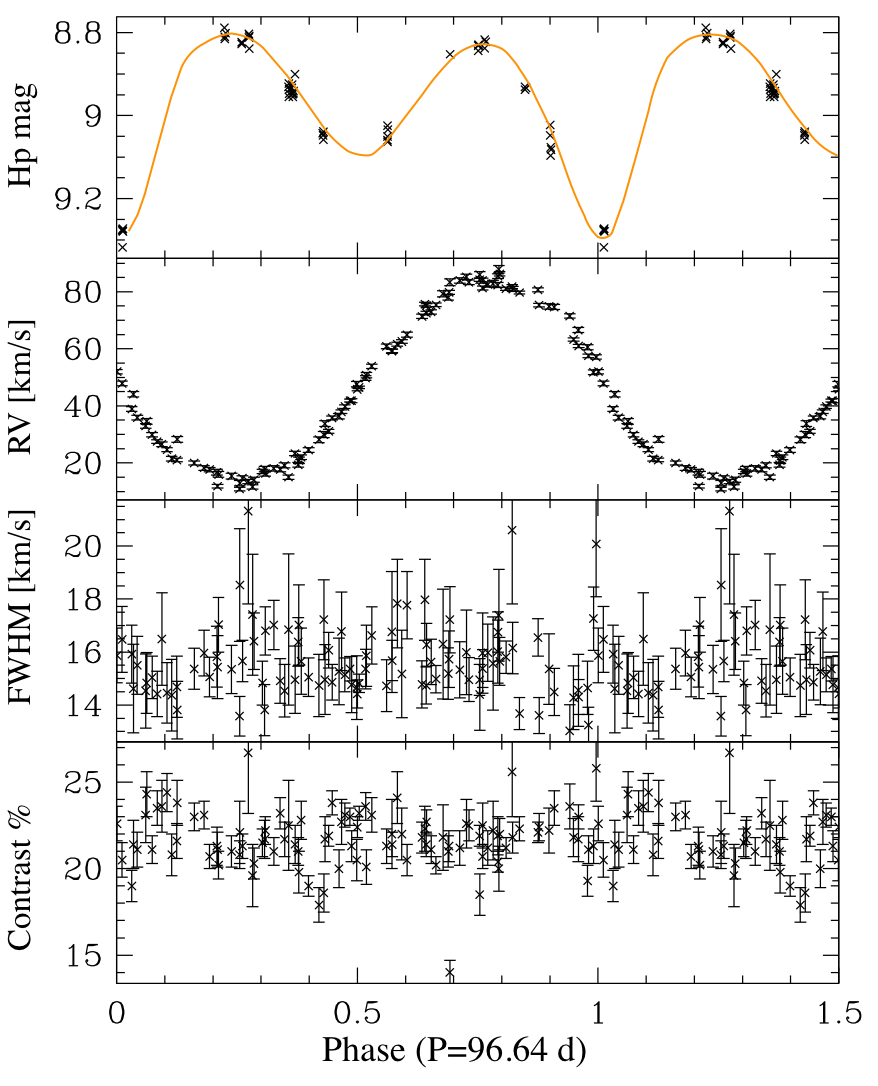}
\caption{The \textit{Hipparcos} photometry (Hp-band data) of an EB eclipsing binary (HIP~39341), together with the CORAVEL parameters derived from the cross-correlation function (CCF): the radial velocity (RV), the full width at half maximum (FWHM), and the CCF contrast.}
\label{fig:EB}       
\end{center}
\end{figure*}

In the case of eclipsing binaries, Fig.~\ref{fig:EB} shows that the eclipses correspond to the mean radial velocity levels (as the velocities of both stars are perpendicular to the line-of-sight), while the variations in FWHM and contrast are not informative, as expected for single-lined spectroscopic binaries (as well as for double-lined eclipsing binaries, when the absorption lines of the two components are disentangled).

The analysis of RV time series and the extraction of their periods will also help to confirm / select the correct orbital period of eclipsing binaries and ellipsoidal variables. In fact, a period search performed on the light curve often leads to half the orbital period, while a period search on the RV curve unambiguously provides the correct orbital period.



We expect the \textit{Gaia} RVS instrument to have similar performance to CORAVEL and significantly improve the characterization and classification of photometric periodic variable stars. 


\section{Conclusion} \label{sec:conclusion}
We have presented several \textit{Gaia} time series related analyses methods, and highlighted several opportunities (and associated difficulties) related to the use of multi-domain time-series measurements. We believe the usage of such rich data will become the norm in astronomy in the coming decades and hence the earlier we start exploring and developing efficient methods to harness the information encoded in it, the better. The \textit{Gaia} data might hence not only spark a revolution in (Galactic) knowledge, but also in the methods that are used to reach these conclusions. 




\bibliographystyle{aasjournal}
\bibliography{bibfileAstro,bibfileOrig} 

\end{document}